\documentclass[a4paper,11pt]{article}
\usepackage{multirow}
\usepackage{graphicx}
\usepackage{pos}

\title{Usage of PEN as self-vetoing structural material in low background experiments}
\author[a]{I.~Abt}
\author[c]{Y.~Efremenko}
\author[b]{M.~Febbraro}
\author[a]{F.~Fischer}
\author[a]{M.~Guitart}
\author[h]{K.~Gusev}
\author[c]{B.~Hackett}
\author[d]{C.~Hayward}
\author[e]{R.~Hod\'ak}
\author[h]{P.~Krause}
\author[a]{B.~Majorovits}
\author*[a]{L.~Manzanillas} 
\author[d]{D.~Muenstermann}
\author[i]{R.~Pjatkan}
\author[f]{M.~Pohl}
\author[f]{R.~Rouhana}
\author[b]{D.~Radford}
\author[e]{E.~Rukhadze}
\author[h]{N.~Rumyantseva}
\author[f]{I.~Schilling}
\author[h]{S.~Schoenert}
\author[a]{O.~Schulz}
\author[h]{M.~Schwarz}
\author[g]{M.~Stommel}
\author[f]{J.~Weingarten}

\affiliation[a]{Max-Planck-Institut  f\"ur  Physik,  80805  Munich,  Germany}
\affiliation[b]{Oak Ridge National Laboratory, Oak Ridge, Tennessee 37830}
\affiliation[c]{Department of Physics and Astronomy, University of Tennessee, Knoxville, Tennessee 37916}
\affiliation[d]{Department of Physics, Lancaster University, Lancaster}
\affiliation[e]{Czech Technical University, Institute of Experimental and Applied Physics, CZ-12800 Prague}
\affiliation[f]{Technische Universität Dortmund, Dortmund}
\affiliation[g]{Leibniz-Institut für Polymerforschung Dresden e.V., 01069 Dresden, Germany}
\affiliation[h]{Physik Department, Technische Universität, München}
\affiliation[i]{Nuvia a.s., 67401 Třebíč, Czech Republic}

\emailAdd{manzanil@mpp.mpg.de}

\abstract{PEN is an industrial polyester plastic which has become interesting for the physics community as a new type of plastic scintillator. PEN scintillates in the blue regime, which is ideal for most photosensor devices. In addition, PEN has excellent mechanical properties and very good radiopurity has been achieved. Thus, it is an ideal candidate for active structural components in low-background experiments. One possible application are holders for germanium detectors operating in cryogenic liquids (LAr, LN2). Such structures can help to reject surface and external backgrounds, boosting the sensitivity of experiments. In this contribution, the R\&D on PEN is outlined and an evaluation of the first production of PEN structures for the LEGEND-200 experiment is reported.}

\FullConference{%
  40th International Conference on High Energy physics - ICHEP2020\\
  July 28 - August 6, 2020\\
  Prague, Czech Republic (virtual meeting)
}

\begin{document}
\maketitle

\section{Introduction}

\begin{figure}[h]
\centering
\begin{minipage}{.3\textwidth}
  \centering
  \includegraphics[trim={0 0 0 0},clip,width=\linewidth]{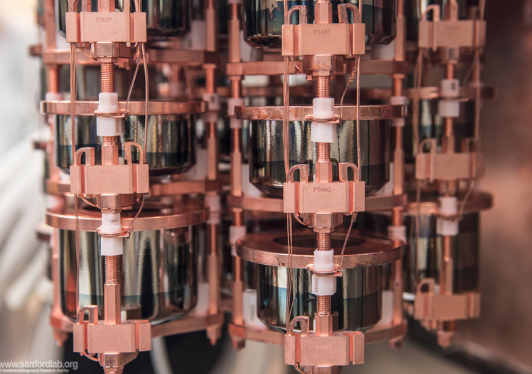}
\end{minipage}%
\begin{minipage}{.33\textwidth}
  \centering
  \includegraphics[trim={0 0 0 0},clip,width=\linewidth]{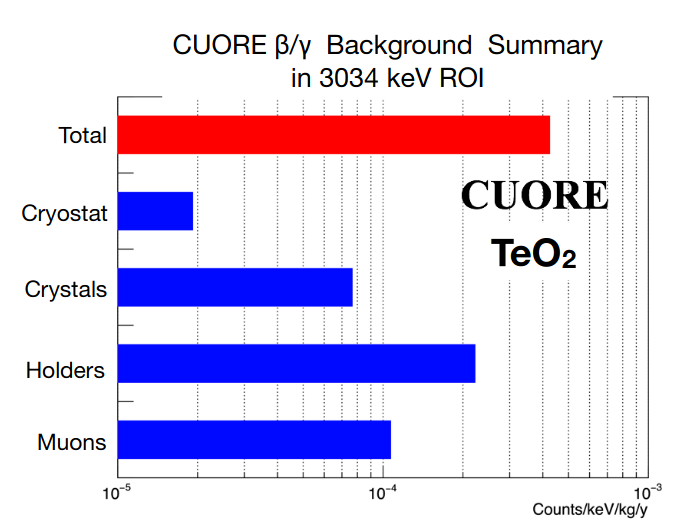}
\end{minipage}
\begin{minipage}{.33\textwidth}
  \centering
  \includegraphics[trim={0 0 0 0},clip,width=\linewidth]{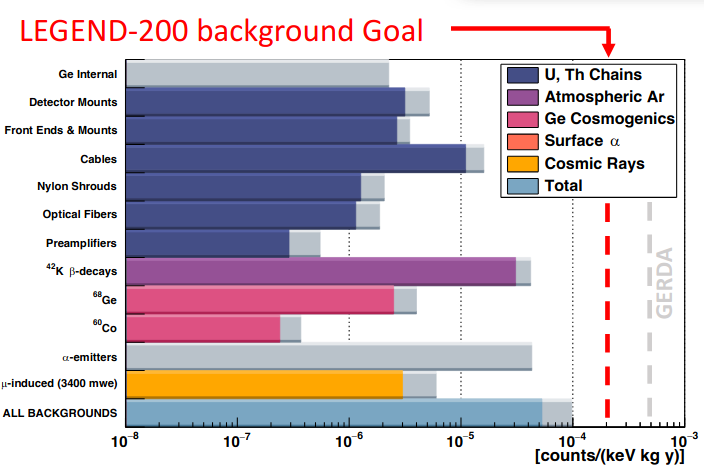}
\end{minipage}
 \caption{Left: Example of inactive copper structures used as support for the Ge detectors in the Majorana experiment \cite{Legend}. Center: Background components in the region of interest in the CUORE experiment, once cuts are applied the remaining background is
dominated by events originating in the support structures \cite{Cuore}.  Right: Expected background composition in the region of interest of the LEGEND-200 experiment \cite{Legend}. An important fraction of background is  expected to be produced by the support structures.}
 \label{fig:bkg1}
\end{figure}

Rare event physics experiments such as dark matter or neutrinoless double beta decay ($0\nu\beta\beta$) searches demand ultra low backgrounds. Hence, ultra pure materials are required for the structural materials and for the detectors itself. An enormous progress has been achieved in the last years on the production of clean materials. However, despite all this progress one major source of background of the current and the next generation of $0\nu\beta\beta$ experiments are the support structural materials (see Figure \ref{fig:bkg1}). 
Some strategies have been adopted to mitigate these external backgrounds. Thus, in some experiments operating at cryogenic temperatures, liquid Argon (LAr) is used to cool down the detectors and at the same time it also serves as an active shielding. Interactions of charged particles in LAr produce VUV light ($\sim127$ nm) \cite{Heindl:2010zz} that can be used to veto external backgrounds. The support structures used to mount the detectors usually consist of optically inactive and nontransparent materials. Besides of being a potential source of backgrounds, structural materials can also absorb light produced by the active shielding, decreasing the efficiency  for background identification in the vicinity of the detectors. In this context, a new approach is under development, and consist in using radio-pure active materials to build the support structures used to mount the detectors. Such materials improve the light collection in the vicinity of the detectors and at the same time have self-vetoing capabilities. This new approach could help to improve the discrimination of surface and external backgrounds increasing the sensitivity of the experiments. 

Poly(Ethylene Naphthalate) known as PEN has been identified as a potential active structural material. PEN is a commercially available polyester, which has a yield strength higher than copper at cryogenic temperatures. In addition, it scintillates in the blue regime, which is ideal for most of photo-sensors.  All these properties make PEN a good candidate to be used as an active structural self-vetoing material \cite{Efremenko:2019xbs}.  

\section{PEN mechanical and optical properties}
PEN [C$_{14}$H$_{10}$O$_{4}$] can be procured as resin or pellets, polymerized slowly, making it easier to shape transparent moulded products without significant crystallization. In order to achieve high quality transparent products, PEN must be cooled from 300~\textdegree~C to 220 \textdegree C in less than 10 seconds during the moulding process. The injection moulding technology in principle allows to produce any arbitrary shape, for example, containers, holding plates, capsules, fibers, among others. PEN moulded shapes have a high structural stability at room and cryogenic temperatures, with a Tensile Strength (209 MPa) higher than copper (100 MPa) at liquid Nitrogen (LN2) temperature (77 K). In addition, PEN has a high chemical resistance  to most acids and organic solvents and therefore can be aggressively cleaned in order to remove  surface contamination. 

\begin{figure}[h]
\centering
\begin{minipage}{.5\textwidth}
  \centering
  \includegraphics[trim=5 8 20 10,clip,width=\linewidth]{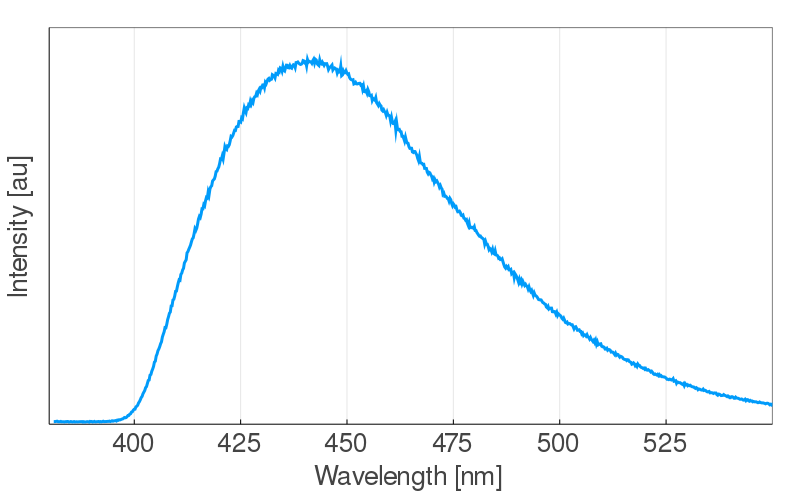}
\end{minipage}%
\begin{minipage}{.5\textwidth}
  \centering
  \includegraphics[trim=5 8 50 20,clip,width=\linewidth]{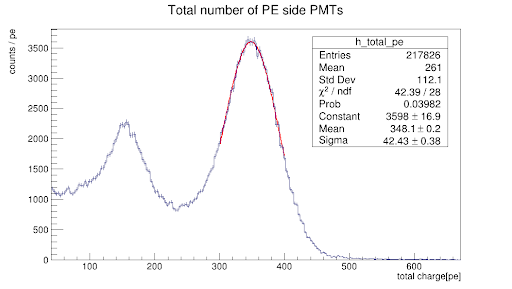}
\end{minipage}
 \caption{Left: Emission spectrum of a PEN sample excited with UV light ($\sim$370 nm). Right: Light output of PEN samples irradiated with a $^{207}$Bi source.}
 \label{fig:PEN_spectrum}
\end{figure}

Transparent PEN moulded shapes scintillate in the blue regime, making it ideal components to be used as active structural materials. The light emission spectrum of a PEN moulded sample is presented in Figure \ref{fig:PEN_spectrum} (Left); it peaks around 445 nm, which is ideal for light detection using standard photo-sensors such as PMTs or SiPMs. 
Figure \ref{fig:PEN_spectrum} (Right) shows the light output of PEN samples excited with a $^{207}$Bi source and measured using PMTs. These first measurements suggest that the scintillation yield of PEN is about 1/3 of standard plastic scintillators. However, PEN has the advantage of providing Pulse Shape Discrimination (PSD),  which allows for alpha decay identification.  Moreover, PEN can shift the  VUV light to visible light. Thus, the VUV light produced by LAr (127 nm) or LXe (175 nm)  can be shifted to about 445 nm with PEN structures with an absolute re-emission probability of about 50\% \cite{Kuzniak:2018dcf}. On the other hand, the light attenuation length of PEN is of the order of few cm, which has a tolerable impact on the light output for structures with a size of the order of a few cm.

\section{PEN application}
One possible application of PEN as an active structural material are holders for germanium detectors, which can be operated in LAr. In this context, using commercially available PEN raw material in form of pellets, PEN tiles of about 1.5 mm thickness have been produced using the injection moulding technology. Before the moulding process, the PEN pellets underwent a systematic cleaning process with the aim of improving the radiopurity of the final moulded products. Finally, the moulding process was realized under clean room conditions. Samples of this production were screened at the LNGS and the LSM underground laboratories. 
The final radiopurity results are presented in Table \ref{tab:PEN:rp}. These radiopurity results meet the strong requirements for materials to be used as support structures in the LEGEND-200 $0\nu\beta\beta$ experiment. These measured PEN tiles will be used to produce holders for the LEGEND-200 experiment.
\begin{table} [h]
\begin{center}
\begin{tabular}{ |c|c r|c c|c| }
 \hline
 \multirow{ 4}{*}{Isotope} & \multicolumn{2}{c|}{Radiopurity  }  & \multicolumn{2}{c|} {Radiopurity  } & Expected  \\ 
         & \multicolumn{2}{c|}{ GeMPI4 at LNGS   }  & \multicolumn{2}{c|} {OBELIX at LSM}  & radiopurity  \\ \cline{2-5}
         & \multicolumn{2}{c|}{ 14.315 kg    }  & \multicolumn{2}{c|} {5.231 kg } &  PEN  holder \\ \cline{2-5} 
         & \multicolumn{2}{c|}{ t = 5851612 s   }  & \multicolumn{2}{c|} {t = 6825600 s  } &   5.3 g \\   \hline \hline
 $^{228}$Ra & \multicolumn{2}{c|}{ $92 \pm 25$\,$\mu$Bq/kg } & \multicolumn{2}{c|}{ $107 \pm 38$\,$\mu$Bq/kg} & $\sim$0.5 $\mu$Bq/piece \\ \hline
 $^{228}$Th & \multicolumn{2}{c|}{ $32 \pm 16$\,$\mu$Bq/kg } & \multicolumn{2}{c|}{ $67 \pm 18$\,$\mu$Bq/kg } & $\sim$0.2 $\mu$Bq/piece \\ \hline
 $^{226}$Ra & \multicolumn{2}{c|}{ $60 \pm 15$\,$\mu$Bq/kg } & \multicolumn{2}{c|}{ $76 \pm 22$\,$\mu$Bq/kg } & $\sim$0.3 $\mu$Bq/piece \\ \hline \hline

\end{tabular}
\end{center}
\caption{Radiopurity achieved for moulded samples using commercially available PEN pellets. The pellets underwent a cleaning process before the injection moulding process. 
  }
\label{tab:PEN:rp}
\end{table}

\begin{figure}[h]
\centering
\begin{minipage}{.5\textwidth}
  \centering
  \includegraphics[trim={5cm 0cm 2cm 0cm},clip,width=0.9\linewidth]{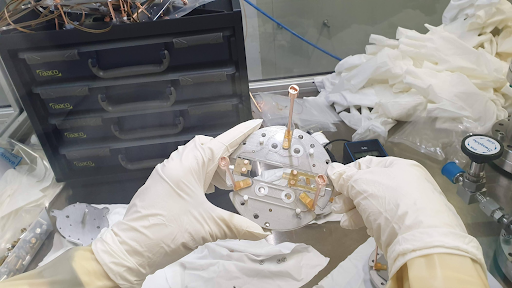}
  \label{fig:test1}
\end{minipage}%
\begin{minipage}{.5\textwidth}
  \centering
  \includegraphics[trim={1cm 2cm 3cm 11cm},clip,width=0.9\linewidth]{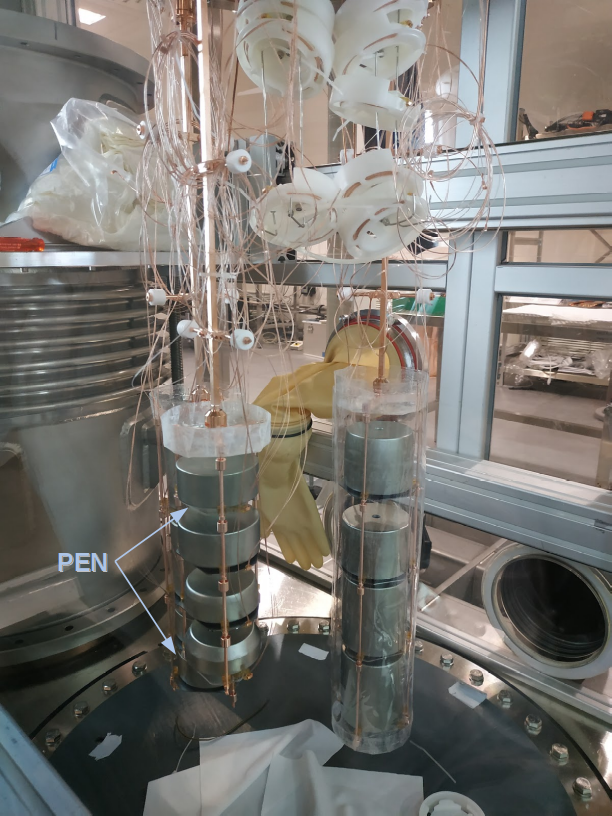}
  \label{fig:test2}
\end{minipage}
 \caption{Left: Low mass PEN holder with optimized geometry on the germanium detector mounting jig while being mounted. Right: Ge detectors mounted with structures consisting of copper, Si and PEN base holders.}
 \label{fig:deployment}
\end{figure}

In order to minimize the mass of the PEN holders that will be used to mount the Ge detectors in the LEGEND-200 experiment, mechanical simulations were used to optimize its design. An example of a holder with optimized geometry on the germanium detector mounting jig is shown in Figure \ref{fig:deployment} (Left). These  PEN holders were tested and validated using the LAr test stand facility of the Technische Universität München (TUM), where a first deployment with HPGe detectors was performed. This study  demonstrated that no significant increase of leakage current is produced due to the usage of PEN holders. Later, additional PEN holders were CNC (Computer Numerical Control) machined at the Physics machine shop of the University of Tennessee under almost clean conditions. These holders were deployed in detector mount prototypes during the LEGEND-200 prototyping tests at LNGS. This experience was used to develop a protocol that will be followed during the production of  PEN holders that will be mounted in the LEGEND-200 experiment.  

PEN samples from the LEGEND-200 production are presently being optically characterized at the  Max-Planck-Institut für Physik in Munich. To this end, different setups consisting of a spectrometer and several PMTs are being used. These setups allow to characterize PEN samples of different dimensions as well as the LEGEND-200 holders for their optical properties. Using these measurements, first estimations of light yield and attenuation of the LEGEND-200 PEN samples have been obtained. Thus, a direct comparison of the light output of PEN samples with polystyrene (SP32) and PVT (EJ-200) samples of the same dimensions demonstrated that the light yield of PEN is at least 3500 photons per MeV. A complete Geant4 optical model of the setups is being developed and will be used in conjunction with more measurements to assess the absolute optical parameters such as light yield, absorption as function of wavelength, emission spectrum, and surface effects. These results will be used as input to the Geant4 LEGEND-200 framework in order to estimate the background rejection efficiency of PEN in the LEGEND-200 experiment.

\section{Conclusions and perspectives}
PEN is an attractive scintillator to be used as active structural material in low background experiments. A 
successful production of low background PEN tiles has been achieved. These tiles are being used to machine low background PEN holders for application in the LEGEND-200 experiment. The preliminary optical results lead to a light yield larger than 3500 photons/MeV, which will allow for self-vetoing capabilities.
Detailed Geant4 simulations of the setups used for the optical characterization of PEN are under development. These studies will provide more precise results of all optical parameters. Finally, further R\&D is being carried out for a potential application of PEN in the LEGEND-1000 experiment.

\end{document}